\documentclass[sii]{ipart_noname}

\usepackage[numbers]{natbib}
\usepackage{mathptmx}
\usepackage{amsmath}
\usepackage{graphics}
\usepackage{graphicx}

\pubyear{2013}
\volume{0}
\issue{0}
\firstpage{1}
\lastpage{1}
\startlocaldefs

\DeclareMathOperator*{\F}{F}
\endlocaldefs

\begin{document}

\begin{frontmatter}

% "Title of the Paper"
\title{motifDiverge: a model for assessing the statistical significance of gene regulatory motif divergence between two DNA sequences}
%\thankstext{t1}{???}
\runtitle{Regulatory motif divergence}

% indicate corresponding author with \corref{}
% \author{\fnms{John} \snm{Smith}\thanksref{t2}\corref{}\ead[label=e1]{smith@foo.com}\ead[label=e2,url]{www.foo.com}}
% \thankstext{t2}{Thanks to somebody}
% \address{line 1\\ line 2\\ \printead{e1}\\ \printead{e2}}

\begin{aug}
\author{\fnms{Dennis} \snm{Kostka},
\ead[label=e1]{kostka@pitt.edu}}
\address{Department of Developmental Biology \\
Department of Computational \& Systems Biology \\
University of Pittsburgh School of Medicine \\
530 45th Street\\
Pittsburgh, PA 15201 \\
\printead{e1}}
\author{\fnms{Tara} \snm{Friedrich},
\ead[label=e2]{tara.friedrich@gladstone.ucsf.edu}}
\address{Gladstone Institutes \\
Integrative Program in Quantitative Biology \\
University of California \\
1650 Owens Street \\
San Francisco, CA 94158 \\
\printead{e2}}
\author{\fnms{Alisha K.} \snm{Holloway},
\ead[label=e3]{alisha.holloway@gladstone.ucsf.edu}}
\address{Gladstone Institutes \\
Division of Biostatistics \\
University of California \\
1650 Owens Street \\
San Francisco, CA 94158 \\
\printead{e3}}
\and
\author{\fnms{Katherine S.} \snm{Pollard}\corref{}
\ead[label=e4]{kpollard@gladstone.ucsf.edu}}
\address{Gladstone Institutes \\
Institute for Human Genetics \\
Division of Biostatistics \\
University of California \\
1650 Owens Street \\
San Francisco, CA 94158 \\
\printead{e4}}

\runauthor{D. Kostka et al.}

\affiliation{University of Pittsburgh, Gladstone Institutes, and University of California, San Francisco}

\end{aug}

\begin{abstract}
Next-generation sequencing technology enables the identification of thousands of gene regulatory sequences in many cell types and organisms. We consider the problem of testing if two such sequences differ in their number of binding site motifs for a given transcription factor (TF) protein. 
Binding site motifs impart regulatory function by providing TFs the opportunity to bind to genomic elements and thereby affect the expression of nearby genes. 
Evolutionary changes to such functional DNA are hypothesized to be major contributors to phenotypic diversity within and between species; but
despite the importance of TF motifs for gene expression, no method exists to test for motif loss or gain. Assuming that motif counts are Binomially distributed, and allowing for dependencies between motif instances in evolutionarily related sequences, we derive the probability mass function of the difference in motif counts between two nucleotide sequences. We provide a method to numerically estimate this distribution from genomic data and show through simulations that our estimator is accurate. Finally, we introduce the R package {\tt motifDiverge} that implements our methodology and illustrate its application to gene regulatory enhancers identified by a mouse developmental time course experiment. While this study was motivated by analysis of regulatory motifs, our results can be applied to any problem involving two correlated Bernoulli trials. 
\end{abstract}

\begin{keyword}
\kwd{testing}
\kwd{gene regulation}
\kwd{motif}
\kwd{ChIP-seq}
\kwd{binomial}
\kwd{transcription factor}
\kwd{regulatory evolution}
\end{keyword}

\end{frontmatter}

%=====================
\section{Introduction}
%=====================

Next-generation sequencing increasingly provides insight into the locations of regulatory regions in the genomes of many organisms, and it gives information about the cell types and developmental stages in which these regulatory elements are active \cite{rivera_2013_mapping}. RNA sequencing (RNA-seq, \cite{mcgettigan_2013_transcriptomics,oszolak_2011_rna}) enables accurate quantification of gene expression, and techniques such as DNase sequencing (DNase-seq, \cite{john_2013_genome-scale}) and Formaldehyde-Assisted Isolation of Regulatory Elements (FAIRE-seq, \cite{giresi_2007_faire}) pinpoint which parts of a genome are in open chromatin and therefore may be associated with regulatory activity in a given cell type. These methods can be coupled with chromatin immunoprecipitation followed by sequencing (ChIP-seq, \cite{furey_2012_chip-seq}) for histone modifications, transcription factors (TFs) and co-factors to further refine predictions of regulatory elements, such as promoters, enhancers, repressors, and insulators \cite{wang_2013_computational}.
Gene expression levels are different between cell types and dynamic during development as the result of regulatory elements that are specifically active in some cells but not in others \cite{laat_2013_topology,wamstad_2012_dynamic}. Therefore, identification of functional regulatory elements and the TFs that recognize them is a key step to characterizing any type of cell. This information also sheds light on transitions between different cell types, such as in the progression to cancer or during cellular differentiation. 

Regulatory genomic elements typically contain multiple motifs for one or more TFs. The TF proteins bind to these motif sequences to combinatorially modulate the expression of nearby genes \cite{maston_2012_characterization}. TF motifs are to some extent degenerate (i.e.,~mutations away from the consensus sequence are tolerated), and therefore they are typically represented as probability distributions over nucleotides ($A$, $C$, $G$, and $T$) at each position in the motif \cite{stromo_2000_dna}. For each TF, this distribution can be represented as position specific probability matrix (PSPM). While TF binding depends on more than just the target DNA sequence (TF concentration, open chromatin, etc.), and even though the binding affinity of a TF towards a stretch of nucleotides is quantitative rather than binary, the presence or absence of TF motifs can be represented as a binary event by scoring how well a sequence matches a TF's PSPM (details below). Because sequence changes can alter how well DNA matches a PSPM, mutations and substitutions can create or destroy motif instances. It is challenging to predict the effect of a single motif loss or gain on the function of a regulatory region, because a loss may be compensated for by a nearby gain. However, a large cumulative change in the number of motifs across a regulatory region can alter expression of nearby genes,
 potentially resulting in differences in organismal traits, such as disease susceptibility. 

To the best of our knowledge there are no existing methods for quantifying divergence between DNA sequences based on differences in motif counts. The primary challenge is that in most biologically meaningful settings the sequences are related through evolution (i.e., they are homologous), and therefore motif instances are correlated. This is the setting we address in this paper: We derive the joint distribution of the number of motifs in the two sequences, and the marginal distribution of the difference in numbers of motifs between the two sequences. From the latter distribution, we show how $p$--values can be computed for testing the null hypothesis of no systematic difference in motif counts between two sequences. We validate our methodology through simulations and apply it to ChIP-seq and RNA-seq data from a developmental time course. 

%================================================
\section{A Model for Regulatory Motif Divergence}
%================================================

\label{sec:modelIntro}
We propose a probabilistic model and test for assessing the statistical significance of the difference in number of motifs for a single TF between two DNA sequences. While the core of our approach is independent of the specifics regarding TF motif modeling, we also provide methodology to estimate the distribution of our test statistic for any TF that has a motif model in the form of a PSPM. 
The sequences may be homologous or not, because our approach does not require (but can make use of) a sequence alignment. In both cases, the two sequences can be short sequence elements (e.g., gene promoter) or concatenations of multiple short sequence elements that share some property (e.g., promoters of multiple genes). For the non-homologous case, any two sequences or sets of sequences can be compared. For example, one might be interested in TFs with significantly different numbers of motifs in promoters of genes that are up-regulated versus down-regulated in a cancer RNA-seq experiment or in comparing gene promoters versus distal enhancers. For the homologous case, one might compare two genotypes present within a single species, such as a disease-associated versus healthy genotype of a gene promoter. The homologous case can also be used across species, for instance, to quantify the regulatory divergence of pairs of homologous regulatory sequences identified via ChIP-seq. We recently took this approach to compare human and fish developmental gene regulation, and we showed that TF motif differences capture functional changes in enhancer sequences better than do standard measures of sequence divergence \cite{ritter_2010_importance}.  

%--------------------------------------------
\subsection{Background: Predicting TF motifs}
%--------------------------------------------

\label{sec:motifpred}
A typical approach to identify TF motifs in  DNA sequences is to scan a sequence one position at a time using a PSPM and predict a motif at any position where the likelihood of a motif-length sub-sequence under the PSPM model is significantly higher than under a background distribution (see below for details) \cite{rahmann_2003_power}. In this context, the PSPM and background distribution are thought of as generative models. Let $M$ be a PSPM of length $l$ (typically about $7$ to $10$bp) over the DNA nucleotide alphabet $\{A,C,G,T\}$, where $M_{ij}$ is the probability of observing nucleotide $i$ at position $j$ in the motif. Let $B_i$ be the probability of observing nucleotide $i$ (at any position) under a background model. Such a background model can, for example, be estimated from the whole genome or from any reasonably long sequence from the species of interest.  Then $L_{ij} := \log(M_{ij}/B_i)$ is the log odds for nucleotide $i$ at position $j$ and $T(x)=\sum_{j=1}^l L_{x_j}$  is the log odds score for a sequence $x=x[1,\ldots,l]$ of length $l$. The distribution of $T$ can obtained numerically, and a log odds score threshold for predicting motif instances can be found in such a way that Type I error, Type II error, or a balance between the two (balanced cutoff) are controlled \cite{rahmann_2003_power}. Alternatives to Type I error control are commonly employed, because false negatives can be important in this application; TFs frequently bind to sequences that are weak matches to their motif (i.e., would be missed with strict Type I error control), and in some cases this weak binding is functional. 

We note that PSPM based log odds scores do not account for dependencies between motif positions, despite the fact that these are known to exist for TF motifs. More sophisticated methods for motif annotation that take relationships between nucleotide positions into account have been developed \cite{siddharthan_2010_dinucleotide,zhao_2012_imporved,mathelier_2013_next}. However, standard PSPM scoring is commonly used, computationally convenient, and has recently been observed to perform well \cite{weirauch_2013_evaluation}. The model we describe in this paper can in principle be applied together with any method for motif prediction.

To scan a sequence $x$ of length $k \ge l$ for motifs, a sliding window approach is typically used. Starting at the first nucleotide $x_1$, compute $T(x_{1\rightarrow l}):=T(x[1,\ldots, l])$. Then, slide the window one nucleotide to start at position $x_2$ and compute $T(x_{2\rightarrow l+1})$. Continue computing $T(x_{i\rightarrow i+l-1})$ until the last test statistic $T(x_{k-l+1\rightarrow k})$ is computed. A motif is predicted at position $i$ if $T(x_{i\rightarrow i+l-1}) > t$ for a log odds score threshold $t$ (see above). Note that subsequent test statistics are not independent, because their underlying sequences overlap. This ``in-sequence'' dependency is often not accounted for, but there are methods that take it into account \cite{pape_2008_compound}. Our model does not explicitly include in-sequence dependency. However, we show that there is a relationship between in-sequence dependence and the dependence between motif counts in two homologous sequences (Appendix). Because of this relationship, our model is able to indirectly account for some in-sequence dependence via its parameter for between-sequence correlation.

%---------------------------------------------------------------------------------
\subsection{Modeling differences in the number of TF motifs between two sequences}
%---------------------------------------------------------------------------------

\label{sec:modelBasic}

Consider two sequences $x$ and $y$ of lengths $k_x$ and $k_y$ (possibly not equal). For a given TF, let a random variable $X_i$ be the indicator for the presence of a motif at position $i$ in $x$, and let $Y_i$ be the corresponding random variable for $y$. We assume the prediction of a motif in a sequence is the result of a Bernoulli trial with a homogeneous success probability along the sequence. Then, the joint distribution of $(X_i,Y_i)$ does not depend on $i$. Next, we define random variables $N_x=\sum_i X_i$ and $N_y=\sum_i Y_i$ for the total number of motifs in each sequence. Marginally $N_x$ and $N_y$ have Binomial distributions. However, note that $X_i$ and $Y_i$ (and therefore $N_x$ and $N_y$) are not necessarily independent, because the sequences $x$ and $y$ are potentially related, for example due to sequence homology or shared regulatory constraints. 
The problem we address here is how to define and estimate the distribution of the difference in the number of motifs between the two sequences $N_{xy}=N_x-N_y$ under dependence of $X_i$ and $Y_i$. Our approach is based on the two underlying, correlated Binomial trials. 

\subsubsection{Equal length sequences}
%-------------------------------------

First consider the case of equal length sequences ($k:=k_x=k_y$), which simplifies the model because there is a corresponding Bernoulli trial in $x$ for each trial in $y$. Let $N_{10}$ be the number of pairs $(X_i, Y_i)$ with $X_i=1$ and $Y_i=0$, and let $N_{01}$ be the number of pairs with $X_i=0$ and $Y_i=1$. 
Then $N_{xy}=N_{10}-N_{01}$. To derive the distribution of $N_{xy}$, we first consider the joint distribution of $N_{10}$ and $N_{01}$, which is multinomial: 
\begin{equation}
\begin{aligned}
\label{eq:nplusminus}
&P(N_{10}=n_{10},N_{01}=n_{01}) = \\
&(n_{10},n_{01}, n-n_{10}-n_{01})!\; p_{10}^{n_{10}}\,p_{01}^{n_{01}}\,(1-p_{10}-p_{01})^{n-n_{10}-n_{01}},
\end{aligned}
\end{equation}
\noindent where 
$(\cdot,\cdot,\cdot)!$ is the multinomial coefficient, $n=k-l+1$ is the number of windows tested for a motif of length $l$, $p_{00}=Pr(X_i=0, Y_i=0)$, $p_{01}=Pr(X_i=0,Y_i=1)$, and so on. 

Notably the joint distribution of $(N_{10},N_{01})$ is independent of $p_{00}$ and $p_{11}$ and only depends on the probabilities for a motif in one sequence and not the other: $p_{01}$ and $p_{10}$. Because $n_{xy} = n_{10} - n_{01}$ can be realized in $\lfloor\frac{n-n_{xy}}{2} \rfloor$ different ways, the distribution of $N_{xy}$ is:
\begin{equation}
\begin{aligned}
\label{eq:eqlendist}
&P(N_{xy}=n_{xy}) = &   \\
& \left\{
\begin{array}{l l}
\displaystyle\sum_{j=0}^{\lfloor \frac{n-n_{xy}}{2} \rfloor} P(N_{10}=n_{xy}+j,N_{01}=j) & \mbox{for } n_{xy} \geq 0\\
\displaystyle\sum_{j=1}^{\lfloor \frac{n-n_{xy}}{2} \rfloor} P(N_{10}=j,N_{01}=|n_{xy}|+j)& \mbox{for } n_{xy} < 0. \\
\end{array}
\right.
\end{aligned}
\end{equation}
Identifying the sums in Equation (\ref{eq:eqlendist}) as hypergeometric series (Appendix), we can rewrite them in terms of the Gaussian hypergemoetric function $ \sideset{_2}{_1}\F$ \cite{petkosevic_1996_ABbook}:
\begin{equation}
\label{eq:hypergeo}
\begin{aligned}
&\sum_{j=0}^{\lfloor \frac{n-n_{xy}}{2} \rfloor} P(N_{10}=n_{xy}+j,N_{01}=j) =& \\
&{n\choose n_{xy}} p_{10}^{n_{xy}} (1-p_{10}-p_{01})^{n-n_{xy}} \;\times& \\ 
&\sideset{_2}{_1}\F\left(\frac{n_{xy}-n}{2} ;\frac{n_{xy}+1-n}{2};n_{xy}+1; \frac{4p_{10}p_{01}}{(1-p_{10}-p_{01})^2}\right),
\end{aligned}
\end{equation}
\noindent with similar results for the other sum. 
Since  $\sideset{_2}{_1}\F(a;b;c;0)=1$, $N_{xy}$ follows a Binomial distribution with parameters $p_{10}$ and $n$ when $p_{01} \rightarrow 0$ . This is as expected, because in this case $P(N_{10}=n_{10},N_{01}=0)$ is Binomial, and there is only one term contributing to the sums in Equation (\ref{eq:eqlendist}). Similarly, for $p_{10} \rightarrow 0$ the distribution $P(N_{10}=0,N_{01}=n_{01})$ is a Binomial with parameters $p_{01}$ and $n$, and $N_{xy}$ has the same Binomial distribution, just mirrored at $n_{xy}=0$.

Finally, we can obtain the mean and variance of $N_{xy}$ from the multinomial distribution of $N_{10}$ and $N_{01}$ (Equation (\ref{eq:nplusminus})):
\begin{equation}
\label{eq:mean_var_k}
\begin{aligned}
&\mbox{E}[N_{xy}]   &=&\;\, n(p_{01}-p_{10}) \\
&\mbox{Var}[N_{xy}] &=&\;\, n \left( p_{10}(1-p_{10}) + p_{01}(1-p_{01}) + 2p_{10}p_{01}  \right).
\end{aligned}
\end{equation}

\subsubsection{Alternative parametrization}
%------------------------------------------

Instead of the parameters $(p_{11},p_{10},p_{01},p_{00})$ we can use the success probabilities of the Bernoulli trials $X_i$ and $Y_i$, plus their correlation. Define $p:= p_{11}+p_{10}$ and $q:= p_{11}+p_{01}$, and let the correlation between the two trials be $\rho := Cov[X_i,Y_i]/\sqrt{Var[X_i]Var[Y_i]}$.
In this parameterization admissible values of $\rho$ depend on $p$ and $q$. Intuitively, it is clear that not all correlation coefficients can be admissible. For instance, if the trials have different success probabilities they cannot at the same time be perfectly correlated. If we assume $0\leq p \leq q \leq \frac{1}{2}$ then $\rho_-\leq \rho \leq \rho_+$ with
\begin{equation}
\begin{aligned}
\rho_- &= -pq / \sqrt{p(1-p)q(1-q)}\\
\rho_+ &= (1-q)p / \sqrt{p(1-p)q(1-q)},\\
\end{aligned}
\end{equation}
so that our model can fully be specified by the success probabilities of the Bernoulli trials and an admissible correlation coefficient. 
We note that the variance of $N_{xy}$ is maximal at $\rho = \rho_-$ (i.e., $p_{11}=0$), not at $\rho=0$ (i.e., $p_{11}=pq$), which corresponds to independent trials. Further, the variance of $N_{xy}$ is minimal at $\rho = \rho_+$ (i.e., $p_{11}=\min(p,q)$). 

\subsubsection{Different length sequences}
%-----------------------------------------

In most situations, even with homologous sequences, the lengths of $x$ and $y$ will not be identical. Suppose without loss of generality that $x$ is the longer sequence so that $k_x \geq k_y$. Our strategy for modifying $P(N_{xy}=n_{xy})$ to account for the length difference is to treat the first $k_y$ nucleotides as in Equation (\ref{eq:eqlendist}) and to derive the distribution for the number of motifs in the remaining nucleotides of $x$. First, note that $N_{xy}=N_1+N_2$, where $N_1$ is a random variable representing the number of motifs in the first $k_y-l+1$ nucleotides of $x$ minus the number of motifs in the corresponding nucleotides of $y$, and $N_2$ represents the number of motifs in the remaining $k_x-k_y$ possible motif start positions in $x$. Then, $N_1$ has the distribution defined in Equation (\ref{eq:eqlendist}) with length parameter $k_y$ (i.e., $n=k_y-l+1$). It is easy to see that $N_2$ only depends on $x$ and is Binomially distributed with success probability $p_{10}+p_{11}$ and $k_x-k_y$ trials, as expected for the remaining Bernoulli trials. 
If $k_y>k_x$, we leave the definition of $N_1$ unchanged, but instead treat the excess trials in $x$ as negative counts of motifs that are subtracted from the count for the same-length segment of length $k_y$. In this case, $N_2$ is Binomially distributed with success probability $p_{01}+p_{11}$ and $k_y-k_x$ trials. Thus, for different length sequences the difference in numbers of motifs is distributed as the convolution of the distributions for $N_1$ and $N_2$:
\begin{equation}
\label{eq:difflength}
\begin{aligned}
&P(N_{xy}=n_{xy}) =&\\
&\left\{ \begin{array}{ll}
	\displaystyle\sum_{j=0}^{k_x-k_y} P_s(N_1=n_{xy}-j)Bin(N_2=j) & \mbox{for} \quad k_x \geq k_y \\
	\displaystyle\sum_{j=0}^{k_y-k_x} P_s(N_1=n_{xy}+j)Bin(N_2=j) & \mbox{for} \quad k_x < k_y, \\
\end{array}
\right.
\end{aligned}
\end{equation}
where $Bin(\cdot)$ is the probability mass function of the Binomial distribution with parameters given above, and $P_s$ denotes the probability mass function of $N_{xy}$ in the case of equal-length sequences (Equation (\ref{eq:eqlendist})). We get the mean and variance of $N_{xy}$ for different length sequences from Equation (\ref{eq:mean_var_k}) and the Binomial distribution:
\begin{gather}
\label{eq:mean_var_kdl}
\begin{aligned}
\mbox{E}[N_{xy}] =\,\,& 
 k_y(p_{10}-p_{01}) + (k_x - k_y)p \\
\mbox{Var}[N_{xy}] =\,\,& 
 k_y  \big( p_{10}(1-p_{10}) + p_{01}(1-p_{01}) + 2p_{10}p_{01}\big)+&\\
	 &(k_x-k_y)p(1-p), \\
\end{aligned}
\raisetag{4\baselineskip}
\end{gather}
where again $k_x\geq k_y$ without loss of generality. Unlike Equation (\ref{eq:eqlendist}), which depends only on $p_{10}$ and $p_{01}$, the distribution of $N_{xy}$ for unequal length sequences (Equation (\ref{eq:difflength})) depends on $p_{11}$ as well (via $p=p_{11}+p_{10}$) and makes full use of the parametrization of $(X_i,Y_i)$.

\subsubsection{Computing $P(N_{xy}=n_{xy})$ and $P(N_{nx} \geq n_{xy})$}
%-----------------------------------------------------------------------

Our main application is to compute a $p$--value for an observed difference in motifs ($n_{xy}$) between two sequences $x$ and $y$. Thus, we are interested in computing a tail probability of the probability mass function of $N_{xy}$ (Equation (\ref{eq:difflength})). To test if $n_{xy}$ is significantly larger compared to what we expect under a null hypothesis we need to  obtain $P(N_{xy}\geq n_{xy})$. Similarly, we need $P(N_{xy}\leq n_{xy})$ to test for significantly fewer motifs in $x$ compared to $y$. 

To numerically evaluate $P(N_{xy}=n_{xy})$, we perform the convolution in Equation (\ref{eq:difflength}) using the fast Fourier transform. A prerequisite for this is the probability mass function $P_s(N_{xy} = n_{xy})$ for the symmetric case ($k_x = k_y$), which we get from Equation (\ref{eq:eqlendist}) and evaluate  up to a pre-specified error $\epsilon \geq 0$. More specifically, let $P_s(N_{xy}=n_{xy}) = \sum_j S_j$, where the summands $S_j$ are taken from Equation (\ref{eq:eqlendist}). Further let $w_j := S_{j+1}/S_j$. Then there exists $j_-$ such that for $j_+$ with $j_-<j_+\leq \lfloor\frac{n-n_{xy}}{2}\rfloor$ (Appendix):
\begin{equation}
\begin{aligned}
\label{eq:directsum}
&P_s(N_{xy}=n_{xy}) = \sum_{j=0}^{j_+}S_j + \epsilon(j_+) \\
&\mbox{with} \quad
0 \leq \; \epsilon(j_+)\;<S_{j_+} \Big( \frac{1-w_{j_+}^{\frac{n-n_{xy}}{2}-j_+}}{1-w_{j_+}}-1\Big).
\end{aligned}
\end{equation}
We evaluate this error bound after each additional term in the sum and stop when a desired precision has been achieved. Additionally, in order to obtain $P_s(N_{xy}=n_{xy})$ for a series of values for $n_{xy}$ the following recurrence relation (Appendix) is useful:
\begin{equation}
\label{eq:recurrence}
\begin{aligned}
&(n-n_{xy})p_{10}P_s(N_{xy}=n_{xy}) = \\
&(1-p_{10}-p_{01})(n_{xy}+1)P_s(N_{xy}=n_{xy}+1)+\\
&p_{01}(n+n_{xy}+2)P_s(n_{xy}+2).
\end{aligned}
\end{equation}
The fast Fourier transform evaluates $P(N_{xy}=n_{xy})$ over an entire range of values for $n_{xy}$, which enables us to compute tail probabilities $P(N_{xy}\geq n_{xy})$, and thereby $p$--values, by direct summation. 

\subsubsection{Estimating model parameters}
%------------------------------------------

\label{sec:modelParameters}

Up to this point, we have treated the model parameters $(p_{10},p_{01},p_{11})$, or alternatively $(p,q,\rho)$, as known. In practice they must be estimated from data before one can compute $p$--values for an observed difference $n_{xy}$ in the number of motif hits between two sequences. The process of predicting TF motifs (Section \ref{sec:motifpred}) suggests several properties that could influence the shape of the probability mass function of $N_{xy}$: 
\newcounter{Lcount}
\begin{list}{($\roman{Lcount})$}{
\usecounter{Lcount}
\leftmargin=2em}

\item \emph{Sequence length.} More predicted motifs can be expected in longer sequences. Also, the larger the length-difference between two sequences, the larger the difference in motifs is expected to be. Both of these effects are explicitly included in our model (via $k_x$ and $k_y$), and we assume that these sequence lengths are known.

\item \emph{Motif information content.} Low information content (i.e., weak or uninformative) PSPMs can lead to more predicted motif instances compared to high information content PSPMs. This effect can be taken into account via the choice of the log odds score threshold $t$ (Section~\ref{sec:motifpred}). For example, selecting a value of $t$ for each TF that controls the Type I error will make motif counts comparable across TFs.

\item \emph{Threshold for predicting motifs.} A loose threshold $t$ for predicting motifs will result in more motif predictions. In our model, the expected number of motifs will be reflected in the parameters $p$ and $q$.

\item \emph{Sequence composition.} For a given background distribution, the probability of a motif prediction will depend on the similarity of the nucleotides favored in the PSPM compared to the nucleotide composition of the sequence. For instance, for a GC-rich motif we expect more motifs in a GC-rich sequence compared to an AT-rich sequence. The parameters $p$ and $q$ account for the sequence composition of $x$ and $y$, respectively. While effects of sequence composition can be further mitigated by using sequence-dependent prediction thresholds $\{t_{xy}\}$ (e.g., corresponding to sequence-dependent background distributions $B_i$), this is not desirable if a consistent threshold is sought for a collection of jointly analyzed sequences.

\item \emph{Relationship of the two sequences.} If the two sequences are homologous, we may expect fewer differences in motifs compared to the case of two independent sequences. As described above, we model the relationship between $x$ and $y$ via a correlation parameter $\rho$, which allows us to accommodate both correlated ($\rho>0$) and uncorrelated ($\rho=0$) sequences.

\end{list} 
Taking these issues into account, we propose the following approaches to parameter estimation. 

{\it Independent sequences: $\quad$} Assume $x$ and $y$ are independent and that motifs are equally likely in both sequences. Then, we can estimate $\hat{p}=\hat{q}:=(n_x+n_y)/(k_x+k_y)$ (which implies $\hat{p}_{10} = \hat{p}_{01}$). With respect to the correlation parameter $\rho$ we have two options. First, we can choose $\hat{\rho}=0$, reflecting the independence of $X_i$ and $Y_i$. In this case, our model is fully specified. An alternative choice for $\hat{\rho}$ is to leverage the relationship between in-sequence dependence (see Section \ref{sec:motifpred}) and the dependence between $x$ and $y$ that is reflected in $\rho$ (Appendix). Specifically, $\lambda_x := P(X_i=1|X_{i-1}=1)$ may be different from $P(X_i=1|X_{i-1}=0)$, and such a correlation ($\lambda_x \neq p$) influences the variance of $N_x$ \cite{klotz_1973_statistical}. % FORMULA 5.5 in KLOTZ
A similar effect holds for $\lambda_y:=P(Y_i=1|Y_{i-1}=1) \neq q$ and the variance of $N_y$. Numerical estimates for $\lambda_x$ and $\lambda_y$ can be obtained (Appendix), and we can then choose an estimator of the between-sequence correlation $\rho$ in such a way that the variance of $N_{xy}$ reflects the in-sequence Markov dependence quantified by the numerical estimates $\hat{\lambda}_x$ and $\hat{\lambda}_y$:
\begin{equation}
\label{eq:insequence}
\hat{\rho} = \frac{-1}{\max(n_x,n_y)} \Big(A(\hat{p},\hat{\lambda}_x,n_x) + A(\hat{q},\hat{\lambda}_y,n_y)\Big),
\end{equation}
where $A(\cdot)$ quantifies the effect of the in-sequence dependence on the variance of $N_x$ and $N_y$ (Appendix, \cite{klotz_1973_statistical}).

{\it Dependent sequences: $\quad$} If $x$ and $y$ are homologous sequences, we propose to estimate model parameters using an evolutionary model that quantifies the probability of nucleotide changes between $x$ and $y$. We will focus on evolutionary models for cross--species data based upon continuous time Markov chains (CTMCs), but population genetics models for genotypes within species could also be used.

Like in the case of independent sequences we estimate $\hat{p}=\hat{q}:=(n_x+n_y)/(k_x+k_y)$. But we estimate the between-sequence correlation $\rho$ via an estimate for $p_{11}$ derived from the evolutionary model. More specifically, suppose there is a motif at position $i$ in $x$ (i.e., $X_i=1$). Consider the probability $p_{1\rightarrow 1}$ that the congruent, homologous sub-sequence of $y$ also contains a motif. We then obtain a numerical estimate $\hat{p}_{1\rightarrow 1}$ based on the sequence composition of $x$ and $y$, an evolutionary model, the PSPM, the background model and the score cutoff $t$ used to predict motifs (Appendix). Finally, an estimate of the probability of a motif in both sequences is $\hat{p}_{11} = \hat{p}\hat{p}_{1\rightarrow 1}$, and the resulting estimator of $\rho$ takes the form:
\begin{equation}
\label{eq:betweensequence}
\hat{\rho} = \big(\hat{p}_{11} - \hat{p}^2) / \big(\hat{p}(1-\hat{p})\big).
\end{equation}
Note that $\hat{\rho} =0$ for independent sequences ($\hat{p}_{11} = \hat{p}^2$), and $\hat{\rho}>0$ for positively correlated sequences $\hat{p}_{1\rightarrow 1} > \hat{p}$. Negative between-sequence correlation is typically not accounted for in evolutionary models, so for homologous sequences we have $\hat{\rho}\geq 0$.

\section{Software package}
\label{sec:software}
We implemented statistical tests for differences in the number of motifs between two sequences in an open source software package, called {\tt motifDiverge}, which is written in the R programming language. The package includes functions for predicting motifs in sequences and computing $p$--values based on an estimate of the distribution of motif differences between two sequences. The difference distribution and $p$--value account for sequence lengths, nucleotide composition of the sequences and the motif, the total number of motifs, and the similarity of the two sequences. The {\tt motifDiverge} package is freely available by request from the first author. 

%=========================
\section{Simulation Study}
%=========================

We performed a study on simulated data to assess whether the model in Equation (\ref{eq:difflength}) describes differences in the number of annotated motifs between two sequences well. In order to assess the model and our proposed heuristics for parameter estimation, we compare the shape of estimated histograms for $P(N_{xy}=n_{xy})$ to the true distribution under different scenarios. We also assess the distribution of $p$--values obtained from data simulated under the null hypothesis. These analyses make use of generative phylogenetic models for pairs of DNA sequences. We simulate independent sequence pairs $(x,y)$, as well as correlated sequences where transitions between corresponding nucleotides in $x$ and $y$ are modeled by a continous time Markov chain (CTMC).

%-------------------------------
\subsection{Simulation Approach}
%-------------------------------
\label{sec:simApproach}

We use a phylogenetic hidden Markov model (phyloHMM) \cite{hubisz_2011_phast} to generate pairs of sequences $(x,y)$. Let $\tau$ denote the evolutionary time separating $x$ and $y$. When $\tau$ is small, $x$ and $y$ are correlated (e.g., homologous), while $\tau \rightarrow\infty$ generates independent sequences. To simulate motif instances, our phyloHMM consists of three states: a background ($BG$) and two motif states ($M_1,M_2$, which are reverse complements of each other). The transition probabilities between these states are $1-\zeta$ for $BG$ to $BG$, $M_1$ to $BG$, or $M_2$ to $BG$, and $\zeta/2$ for $BG$ to $M_1$, $BG$ to $M_2$ or between $M_1$ and $M_2$ (Appendix). The parameter $\zeta$ encodes motif prevalence. The background state consists of a CTMC with a strand-symmetric and time-reversible rate matrix estimated from neutrally evolving sites in primate genomes (46-way Conservation track from the UCSC Genome Browser, {\tt http://} {\tt genome.ucsc.edu}). It emits two corresponding nucleotides (one in sequence $x$ and one in sequence $y$) separated by evolutionary distance $\tau$ (i.e., there are $\tau$ expected substitutions between $x$ and $y$ per nucleotide). The motif state consists of a similar CTMC except that the equilibrium probabilities of each position equal the probability distribution given by the TF's PSPM (or its reverse complement). Each motif state emits two sequences of motif-length (one for $x$ and one for $y$). 

We repeatedly generated sequence pairs $(x,y)$ and predicted motifs for the transcription factor {Nkx2-5} using a log odds score threshold $t$ with a false positive rate (Type I error, see section \ref{sec:motifpred}) for motif hits of 1\%. Sequence pairs were generated with different lengths ($k_x,k_y$), between sequence correlation parameters $\tau$, and motif-prevalence parameters $\zeta$. To simulate $k_x \neq k_y$, we generate two sequences of the longer length and then delete the excess nucleotides from the shorter sequence. In most simulations, the motif prevalence is the same in $x$ and $y$, so that we are simulating data reflecting $P(N_{xy}=n_{xy})$ under the null hypothesis of no motif differences between $x$ and $y$. 

For each simulation scenario, we computed three estimates of $P(N_{xy}= n_{xy})$: maximum likelihood estimation of the full distribution, a Gaussian distribution with mean and variance estimated from the simulated data, and the same Gaussian estimator with continuity correction. We also estimated $p$--values using different estimates for the model parameters (see Section \ref{sec:modelParameters}) and using a counting method (Appendix). 

%------------------------------
\subsection{Simulation Results}   
%------------------------------
\label{sec:simResults}

\begin{figure}
\flushleft
\includegraphics[width=.54\textwidth]{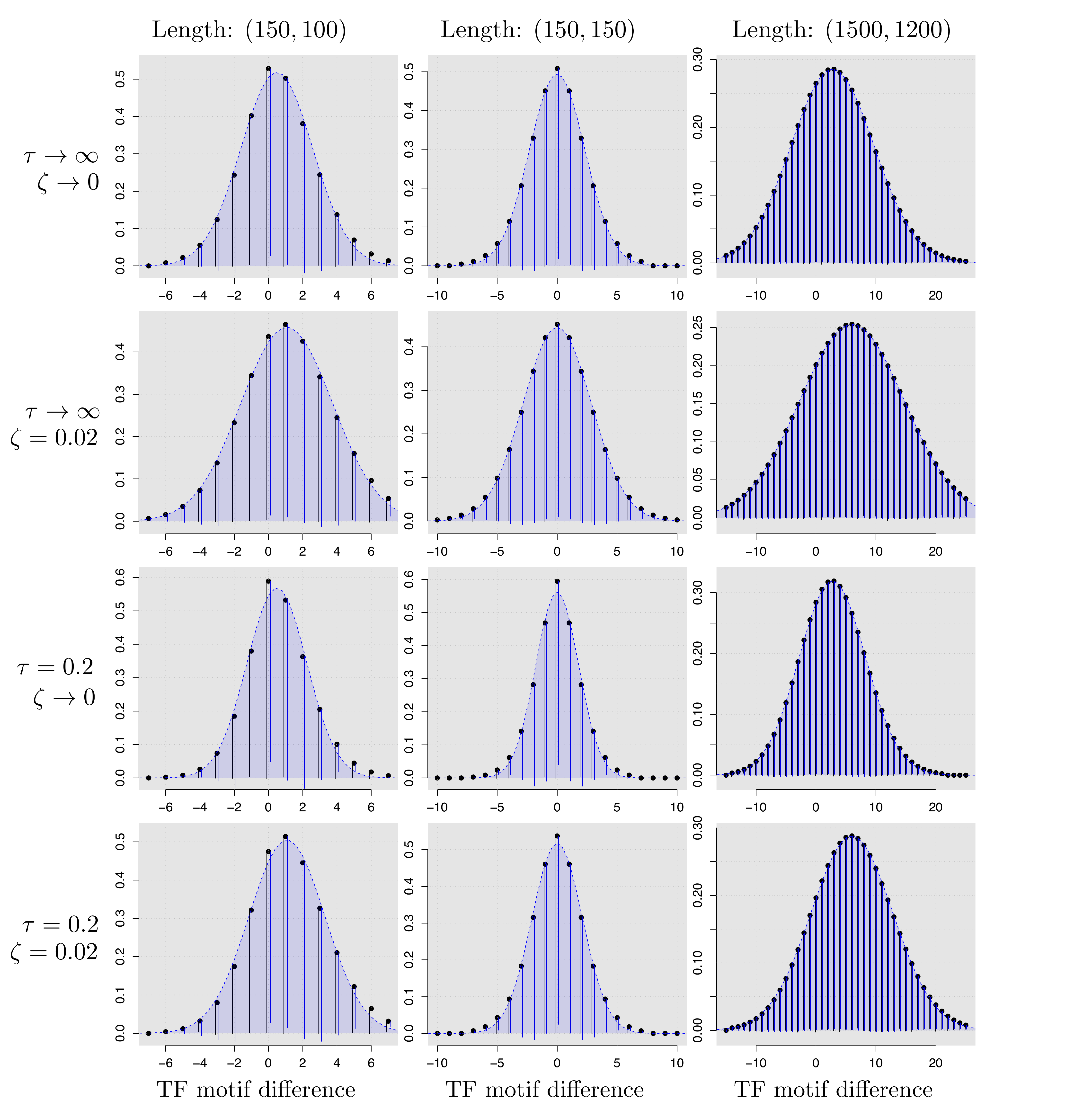}
\caption[]{\label{fig:nkx} $P(N_{xy}=n_{xy})$ describes differences in motif hits well. The rows show different between-sequence dependence, the columns different sequence lengths.}
\end{figure}

First, we show that the proposed estimators of $P(N_{xy}= n_{xy})$ describe differences in motif hits well. Figure \ref{fig:nkx} shows results for three combinations of $(k_x,k_y)$ (columns) and four combinations of ($\tau,\zeta$) (rows). For each scenario, we simulated 100,000 data sets.
Each plot shows a hanging rootogram \citep{tukey_1972_displays} of the differences in the number of observed Nkx2-5 motifs. That is, the vertical axis denotes the square root of the probability, and the horizontal axis the difference in motif counts. The solid circles correspond to the maximum likelihood fit of $P(N_{xy}=n_{xy})$ to the simulated data. The blue dashed lines correspond to a Gaussian approximation with the estimated mean and variance, and the blue vertical bars are the corresponding Gaussian values with continuity correction. These should be compared to the lengths of the black vertical bars, which correspond to the true frequencies of $n_{xy}$ in the simulation. 
The first two rows show simulations for independent sequences ($\tau \rightarrow \infty$) for different values of $\zeta$, while in the second two rows $x$ and $y$ are related ($\tau = 0.2$ expected substitutions per nucleotide). Across these different scenarios, we find that all three estimators of $P(N_{xy} = n_{xy})$ very accurately capture the observed distribution of motif-count differences in our simulations. In other words, the black vertical bars nearly all end at zero; the blue bars are often similar in length to the black bars, and the dotted blue density in general matches the other three distributions fairly closely.  

%---------------------------------------------------------------------------------------
%----------------------------------------------------------------------------------------

\begin{figure}
\includegraphics[width=.49\textwidth]{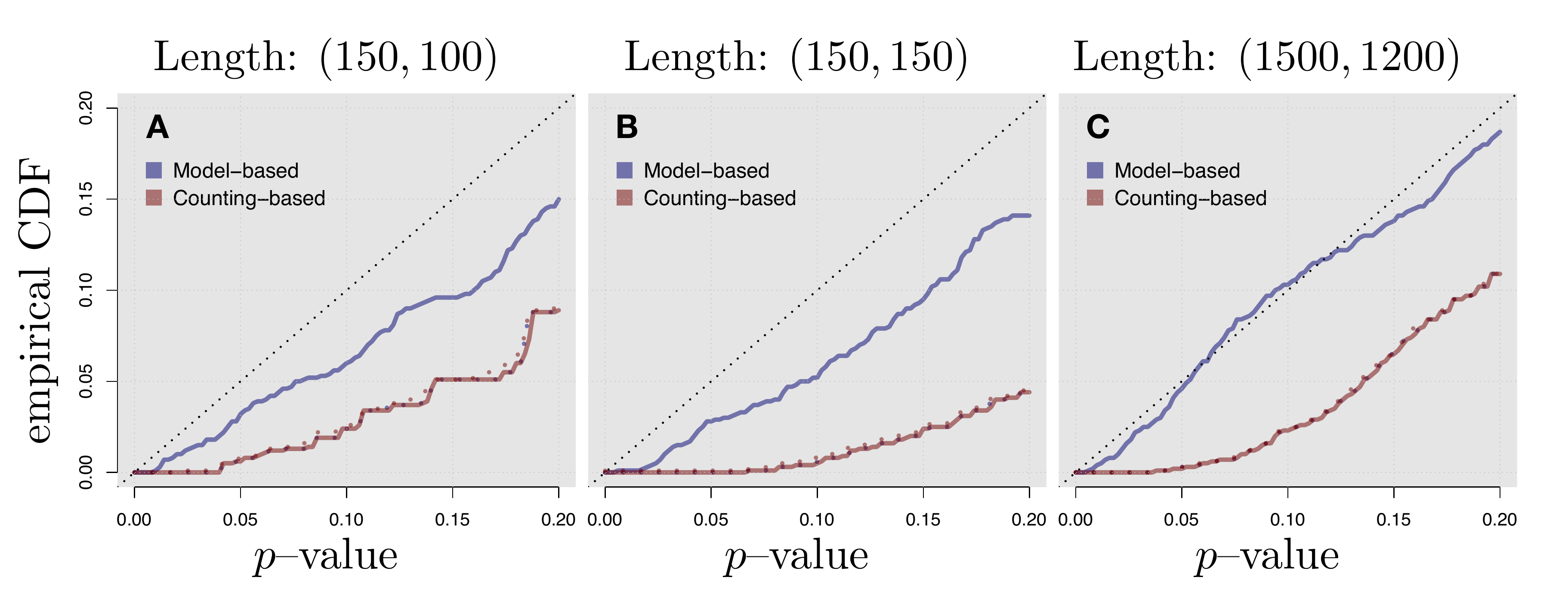}
\caption[]{\label{fig:pvalDist} Partial empirical CDF of 1,000 $p$--values computed using different parameter estimates for data simulated under the null hypothesis. Panels A to C show different sequence lengths.}
\end{figure}

Next, we looked at the accuracy of our estimated $p$--values. We simulated 1,000 sequence pairs with $\tau = 0.02$, $\zeta = 0.02$, and three combinations of sequence lengths $(k_x,k_y)$. Figure \ref{fig:pvalDist} summarizes the results.
Each panel shows the (partial) empirical cumulative distribution function (CDF) of $p$--values obtained from different parameter estimates. The blue lines represent model-based estimates, whereas the red lines represent count-based estimates (see Appendix for definitions of different parameter estimates). The solid lines treat the sequence-pairs as homologous (which is how the data was generated), whereas the dotted lines assume independence between $x$ and $y$. We find that our $p$--values are mostly conservative, and that for longer sequences they become approximately uniformly distributed for smaller $p$. We also see that model-based $p$--values that take between-sequence correlation into account correspond to lower false positive rates compared $p$--values based on other parameter estimates. This implies they appropriately leverage the between-sequence dependence present in the simulated sequence pairs. Interestingly, the estimates assuming uncorrelated sequence pairs are very similar for count-based and for model-based parameter estimates. In light of the greater computational effort for model-based estimates this may suggest the usage of count-based estimates for non-homologous sequences.

%---------------------------------------------------------------------------------------
%----------------------------------------------------------------------------------------

\begin{figure}
\includegraphics[width=.5\textwidth]{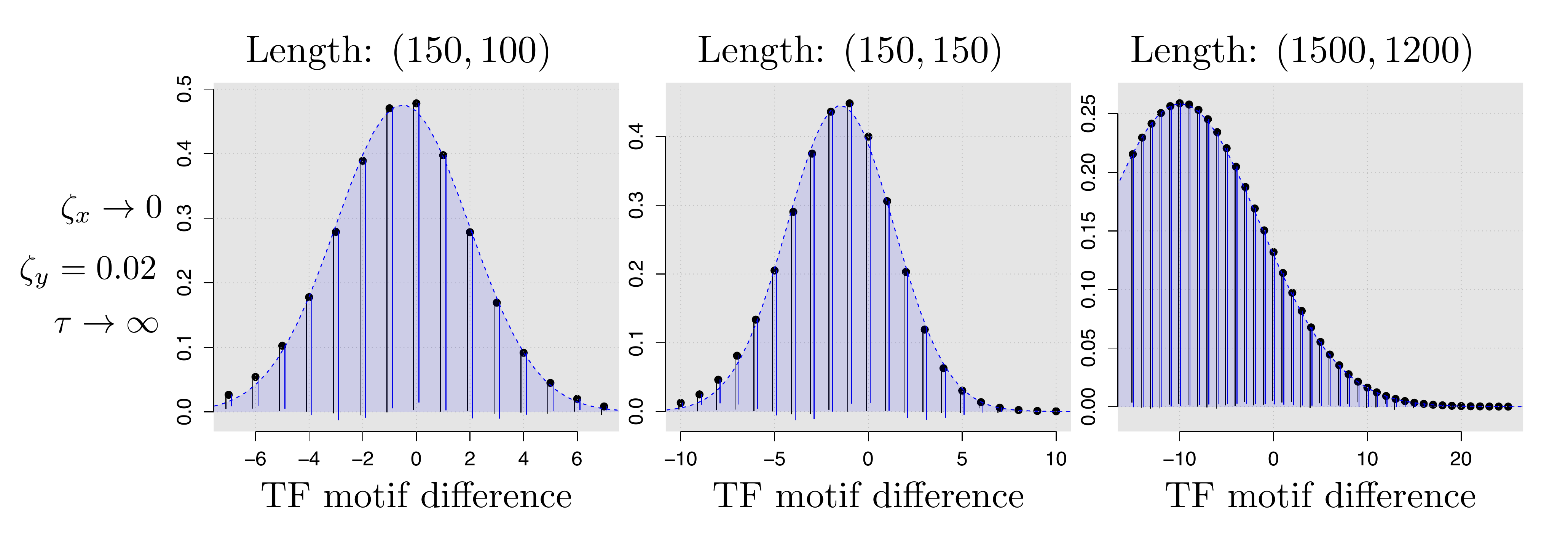}
\caption[]{\label{fig:dzeta} $P(N_{xy}=n_{xy})$ for TF motif differences for sequences with different motif prevalence ($\zeta_x$ vs.~$\zeta_y$). }
\end{figure}

Finally, to assess the model fit of $P(N_{xy} = n_{xy})$ when motif prevalence is different between $x$ and $y$, we simulated 100,000 sequence pairs in the following way. Sequence $x$ was simulated from a phyloHMM with $\zeta_x \rightarrow 0$ and sequence $y$ from a model with $\zeta_y = 0.02$. Taking single sequences from two different phyloHMMs corresponds to $\tau \rightarrow \infty$. Figure \ref{fig:dzeta} is analogous to Figure \ref{fig:nkx} and shows the result. We find that even when motif prevalence is different, our estimators of $P(N_{xy} = n_{xy})$ accurately capture the properties of the true, simulated distribution of $N_{xy}$.

%---------------------------------------------------------------------------------------
%----------------------------------------------------------------------------------------

%=================================================================================
\section{Motif divergence in gene regulatory enhancers during cardiac development}
%=================================================================================

To illustrate the use of {\tt motifDiverge} on genome sequence data, we analyze a collection of gene regulatory elements identified via ChIP-seq for the active enhancer-marking histone modification histone 2 lysine 27 acetylation (H3K27ac) by Wamstad {\it et al.} \cite{wamstad_2012_dynamic}. This study identified genomic sequences marked by H3K27ac in mouse embryonic stem cells (ESCs) and at several subsequent developmental time points along the differentiation of ESCs into cardiomyocytes (CMs), which are beating heart cells. Our analysis uses these cell type specific enhancer sequences to illustrate applications of {\tt motifDiverge} to both non-homologous and homologous sequences. Tissue development is a useful system for illustrating our approach, because active regulatory elements and TFs that are important for regulating gene expression differ across cell types and between species.

%-----------------------------------------------------------------------
\subsection{Motif divergence between mouse and human enhancer sequences}
%-----------------------------------------------------------------------

We first explored the use of {\tt motifDiverge} to quantify motif differences between homologous sequences. For each of the 8,225 H3K27ac-marked enhancers from mouse CMs, we identified the homologous human sequence (if any) using the whole-genome, 100-way vertebrate multiple sequence alignments available from the UCSC Genome Browser ({\tt http://genome.ucsc.edu}), which are based on the hg18 and mm9 genome assemblies. It is interesting to compare CM gene regulation between these two species, because there are a number of structural and electrophysiological differences between their hearts. We identified 1,345 orthologous human-mouse sequence pairs that were at least 20 nucleotides long. For each enhancer pair, we predicted motifs in the human and mouse sequence with JASPAR PSPMs ({\tt http://jaspar.genereg.net}) for all 34 TFs expressed in mouse CMs (fragments per kilobase per million sequenced (FPKM) $>10$) and a log odds score threshold that corresponds to a Type I error rate of $1\%$. Then we tested for TFs with significant differences in motif counts between human and mouse in each CM enhancer region. 

After adjusting for multiple testing using the Benjamini-Hochberg false discovery rate (FDR) controlling procedure \cite{benjamini_1995_fdr}, we found that most enhancers (74\%) show evidence of significant differences in motif counts for at least one TF (FDR$<5\%$). Slightly more than half of CM enhancers (55\%) have significant differences in motif counts for multiple TFs, and several have significant differences for fifteen or more TFs. Conversely, most TFs only have significant differences in counts between human and mouse for a small percentage of CM enhancers. The TFs with the largest percentage of enhancers showing significant differences are listed in Table (\ref{tab:hgmm}). These TFs are promising candidates for understanding differences in CM gene regulation between humans and mice. Interestingly, Sp1 has many enhancers with significantly more motifs in human (19\%) and nearly as many with more motifs in mouse (15\%), suggesting that it may target quite different sets of enhancers--and potentially different genes--in the two species.  

\begin{table}
\begin{tabular}[c]{|r|c|}
\hline
\multicolumn{2}{|c|}{{\em Transcription factors with more motifs in mouse}}\\
\hline
{\bf TF}&{\bf Proportion of CM enhancers}\\
\hline
Prrx2&0.29\\
Cad&0.23\\
Mef2a&0.23\\
Arid3a&0.18\\
Sp1&0.15\\
\hline
\multicolumn{2}{|c|}{{\em Transcription factors with more motifs in human}}\\
\hline
{\bf TF}&{\bf Proportion of CM enhancers}\\
\hline
Sp1&0.19\\
Egr1&0.19\\
Btd&0.12\\
Fhl1&0.083\\
Id1&0.080\\
\hline
\end{tabular}
\caption{Transcription factors with the most enhancers showing significant divergence in motif counts between human and mouse sequences.}
\label{tab:hgmm}
\end{table}

%----------------------------------------------------------------------------------
\subsection{Differences in motifs between enhancers active in different cell types}
%----------------------------------------------------------------------------------

Next, we used {\tt motifDiverge} to compare motif counts between non-homologous sequence pairs. This application also illustrates how {\tt motifDiverge} can be applied to perform a single test to compare two sets of sequences. We concatenated the sequences of the 10,338 H3K27ac-marked regions in CMs to create a single, long sequence containing all the active enhancers for this cell type. Then, we generated a similar concatenation of all 7,162 enhancers from ESCs. Any genome sequence marked by H3K27ac in both ESCs and CMs was removed from both data sets, so that the  resulting two ESC and CM enhancer sequences were non-overlapping. We predicted motifs in the ESC and CM sequences as described above with PSPMs for all 49 TFs expressed in either cell type. Then we tested for TFs with significant differences in motif counts between the combined enhancer regions of the two cell types. At FDR$<5\%$, we found several TFs with significantly different numbers of motifs in ESC versus CM enhancers (Table (\ref{tab:esccm})). 

To better understand the biological meaning of these results, we used RNA-seq data from these two cell types to quantify the expression of each TF. Several TFs are only highly expressed in one cell type. For example, motif count and expression are some times both elevated in one cell type compared to the other. For instance, Cad is more highly expressed and has significantly more motifs in ESCs, suggesting a possibly important role in pluripotency. In other cases, such as Ctcf and Rest, the TF is expressed in both cell types, but at a lower level in the one with more motifs. For these TFs, the larger number of motifs in one cell type may be necessary to compensate for their reduced expression. Finally, RNA-seq data can help us filter out significant motif differences that are not biologically meaningful. For example, Nkx2-5 has significantly more motifs in ESC compared to CM enhancer sequences. However, Nkx2-5 is not expressed in ESCs, making it unlikely that the additional motifs affect ESC gene regulation. Similarly, Pou5f1 (also known as Oct4) has more motifs in CM enhancers but is not expressed in CMs, which make sense since this TF plays an important role in pluripotency ({\tt http://www.genecards.org}). 

These analyses show how {\tt motifDiverge} can be used to analyze data from ChIP-seq experiments and how RNA-seq data can be used to filter and interpret {\tt motifDiverge} findings, leading to robust conclusions about the role of sequence differences in gene regulation. 

\begin{table}
\begin{tabular}[c]{|r|l|l|l|}
\hline
\multicolumn{4}{|c|}{{\em Transcription factors with more motifs in ESC}}\\
\hline
&{\bf FDR adjusted}&{\bf ESC}&{\bf CM}\\
{\bf TF}&{\bf $p$--value}&{\bf Expression}&{\bf Expression}\\
\hline
Arid3a&$<1\mbox{e-}{15}$&4.60&14.16\\
Cad&$<1\mbox{e-}{15}$&74.56&23.44\\
Prrx2&2.4e-10&3.80&33.15\\
Id1&2.3e-9&72.81&70.79\\
Nkx2-5&5.2e-6&0.96&161.63\\
Foxd3&0.021&17.50&0.066\\
\hline
\multicolumn{4}{|c|}{{\em Transcription factors with more motifs in CM}}\\
\hline
&{\bf FDR adjusted}&{\bf ESC}&{\bf CM}\\
{\bf TF}&{\bf $p$--value}&{\bf Expression}&{\bf Expression}\\
\hline
Ctcf&$<1\mbox{e-}{15}$&38.26&13.36\\
Egr1&$<1\mbox{e-}{15}$&17.21&167.44\\
Esrrb&$<1\mbox{e-}{15}$&105.10&0.58\\
Gabpa&$<1\mbox{e-}{15}$&20.43&10.57\\
Klf4&$<1\mbox{e-}{15}$&34.51&5.34\\
Myc&$<1\mbox{e-}{15}$&20.68&	2.47\\
Mycn&$<1\mbox{e-}{15}$&136.69&11.86\\
Nfil3&$<1\mbox{e-}{15}$&2.75&24.077\\
Nfkb1&$<1\mbox{e-}{15}$&9.90&13.93\\
Nfya&$<1\mbox{e-}{15}$&6.99&15.41\\
Pou5f1&$<1\mbox{e-}{15}$&688.11&0.13\\
Rela&$<1\mbox{e-}{15}$&10.15&17.00\\
Rest&$<1\mbox{e-}{15}$&44.21&12.90\\
Rfx1&$<1\mbox{e-}{15}$&13.37&7.59\\
Srf&$<1\mbox{e-}{15}$&21.90&29.67\\
Stat3&$<1\mbox{e-}{15}$&10.34&39.50\\
Tead1&$<1\mbox{e-}{15}$&13.95&25.53\\
Ttk&$<1\mbox{e-}{15}$&18.02&2.13\\
Yap1&$<1\mbox{e-}{15}$&30.55&37.28\\
Zfp423&$<1\mbox{e-}{15}$&13.045&2.50\\
Nfe2l2&$<1\mbox{e-}{15}$&24.40&22.24\\
Fhl1&2.82e-13&30.42&36.011\\
Pbx1&9.61e-11&3.33&22.94\\
E2f1&9.93e-11&21.093&5.48\\
Tbp&1.27e-08&19.075&6.62\\
Usf1&8.52e-08&30.35&19.79\\
Max&6.00e-05&27.013&16.61\\
Irf1&0.00023&20.89&4.25\\
Mef2a&0.00094&2.81&29.53\\
Sp1&0.033&22.83&15.57\\
\hline
\end{tabular}
\caption{Transcription factors with significant differences in TF motif counts between ESCs and CMs. Expression values are fragments per kilobase per million fragments sequenced (FPKM).}
\label{tab:esccm}
\end{table}

%===================
\section{Conclusion}
%===================

In this paper, we propose a new model for the difference in counts between two correlated Bernoulli trials representing numbers of TF motifs in a pair of DNA sequences. Our major results are the model derivation, accurate methods for parameter estimation, and a software package called {\tt motifDiverge} that can be used to predict TF motifs and to perform tests comparing motif counts in two sequences. We illustrate the use of {\tt motifDiverge} to discover TFs with significant differences in motifs ($i$) between two species, or ($ii$) between two cell types. These applications demonstrate the power of our methodology for discovering specific genes and regulatory mechanisms involved in species divergence and tissue development through careful analysis of ChIP-seq data. 

Sequence divergence is usually measured in numbers of DNA substitutions or model-based estimates of rates of substitutions. These measures do not account for whether or not substitutions create or destroy TF motifs and are not well suited to quantify functional divergence \cite{ritter_2010_importance}. Our tests capture how changes to DNA sequences affect their TF motif composition, and therefore they provide a more meaningful measure of divergence for regulatory regions. Hence, our model will be useful for understanding when non-coding mutations affect or do not affect the function of regulatory sequences. This information will enable, for example, identification of causal mutations in genomic regions identified as associated with diseases or other phenotypes. Since the majority of these genome-wide association study (GWAS) hits are outside of protein-coding regions \cite{hindorff_2009_gwas}, {\tt motifDiverge} has the potential to have a large impact on human genetics research. 

In future work, it would be interesting to extend our approach to model the joint distribution of multiple correlated Bernoulli trails and univariate summary statistics (e.g., sums, differences) of this distribution. As with two sequences, the main challenge is modeling correlations between the sequences. The phylogenetic tree models we used here can measure relationships between multiple homologous, but not equally related, DNA sequences; therefore they could provide a natural solution to this problem. 

We focus on comparing counts of TF motifs in two (possibly homologous) sequences, but our model is not specific to motifs in any way. The random variables $N_x$ and $N_y$ could represent other features of interest in two related DNA sequences, such as counts of microRNA binding sites, repetitive elements, polymorphisms, or experimentally measured events (e.g., ChIP-seq peaks). In fact, the two Bernoulli trials do not need to measure events on sequences, and our model could be applied to many other types of correlated count data. 

%%%%%%%%%%%%%%%%%%%%%%%%%%%%%%%%%%%%%%%%%%%
\section*{Appendix}
%%%%%%%%%%%%%%%%%%%%%%%%%%%%%%%%%%%%%%%%%%%
 
\subsection*{Derivation of Equation (\ref{eq:hypergeo})}
%=======================================================

\emph{$P(N_{xy}=n_{xy})$ is a hypergeometric function for $k_x=k_y$}
\vspace{1ex}

\label{sec:ap:hypergeo}

The probability mass function of $N_{xy}$ for equal length sequences (Equation (\ref{eq:eqlendist})) can be written as a sum: $P(N_{xy}=n_{xy}) = \sum_j S_j$, with the summands $S_j$ given by Equation (\ref{eq:nplusminus}). Taking the ratio of two successive summands we get:
\begin{equation}
\label{eq:ap:pk_sumrat}
\begin{aligned}
&{S_{j+1}}/{S_j}= \\
&\frac{(n-n_{xy}-2j)(n-n_{xy}-2j-1)}{(j+n_{xy}+1)(j+1)} \frac{p_{10}p_{01}}{(1-p_{10}-p_{01})^2} = \\
&\frac{(j+\frac{n_{xy}-n}{2})(j+\frac{n_{xy}+1-n}{2})}{(j+n_{xy}+1)(j+1)} \frac{4p_{10}p_{01}}{(1-p_{10}-p_{01})^2}.
\end{aligned}
\end{equation}
We note that this is a rational function in $j$, $n_{xy}$ and $n$ and identifies the arguments $(n_{xy}-n)/2$, $(n_{xy}-n+1)/2$ and $(n_{xy}+1)$ of the Gaussian hypergeometric function in Equation (\ref{eq:hypergeo}) \cite{petkosevic_1996_ABbook}.

\subsection*{Derivation of Equation (\ref{eq:directsum})}
%=======================================================

\emph{Error bound for evaluating $P(N_{xy}= n_{xy})$ for $k_x=k_y$}
\vspace{1ex}

\label{sec:ap:dirsum}

Let $w_j := S_{j+1}/S_{j}$. From Equation (\ref{eq:ap:pk_sumrat}), we get that increasing $j$ decreases the numerator $S_{j+1}$ and increases the denominator $S_j$, so that  $w_j$ is decreasing in $j$. Therefore, there exists $j_-$, with $w_{j_-} < 1$ (i.e., the summands $S_j$ are decreasing for $j\geq i_-$). The error $\epsilon(j_+)$ of truncating the sum over $j$ at $j_+ \geq j_-$ is then:

\begin{equation}
\begin{aligned}
&\epsilon(j_+) = \\
&       \sum_{j=j_++1}^{\lfloor\frac{n-n_{xy}}{2}\rfloor} S_j
        = \sum_{j=j_++1}^{\lfloor\frac{n-n_{xy}}{2}\rfloor} w_{j-1}w_{j-2}\ldots w_{j_+} S_{j_+} \; < \\
        &\sum_{j=1}^{\lfloor\frac{n-n_{xy}}{2}\rfloor -j_+} w_{j_+}^jS_{j_+}
        = \;S_{j_+}\left(\frac{1-w_{j_+}^{\lfloor\frac{n-n_{xy}}{2}\rfloor-j_+}}{1-w_{j_+}} - 1\right),
\end{aligned}
\end{equation}
where we have used the following: ($i$) $S_j = (S_j/S_{j-1})S_{j-1} = w_{j-1}S_{j-1}$, ($ii$) $S_j$ are decreasing for $j\geq j_-$, ($iii$) $S_j \leq 1$ are non-negative multinomial probabilities (see Equation (\ref{eq:eqlendist})), and ($iv$) the geometric sum. Thus, to estimate the probability mass function of $N_{xy}$ to a desired precision $\epsilon$, $\sum_j S_j$ an be truncated at the first $j_+ \geq j_-$ for which $\epsilon(j_+) \leq \epsilon$. 

\subsection*{Derivation of Equation (\ref{eq:recurrence})}
%=========================================================

\emph{Recurrence relation for $P(N_{xy} = n_{xy})$ for $k_x = k_y$}
\vspace{1ex}

\label{sec:ap:rec}

Let $P(N_{xy}=n_{xy}) = \sum_{j} S(j,n_{xy},n)$, where the summands $S(j,n_{xy},n)$ are taken from Equation (\ref{eq:eqlendist}). 
Recurrence relations in $n$ and $n_{xy}$ can be obtained via the Zeilberger algorithm \cite{petkosevic_1996_ABbook}, for instance as implemented in the computer algebra system Maxima ({\tt http://sourceforge.net/projects/maxima}). For a recurrence in $n_{xy}$, the Maxima code is:
\begin{footnotesize}
\begin{verbatim}
(%1) Sj : n!/((n_{xy}+j)!*i!*(n-n_{xy}-2*j)!)
         *p10^(n_{xy}+j)*p01^j*(1-p10-p01)^(n-n_{xy}-2*j) $
(%2) load(zeilberger) $
(%3) Zeilberger(Sj,j,n_{xy});
(%o3) [[-(j*(n+n_{xy}+2)*p10)/(n_{xy}+j+1),
       [(n-n_{xy})*p10,(n_{xy}+1)*(p10+p01-1),
       -(n+n_{xy}+2)*p01]]]
\end{verbatim}
\end{footnotesize}
This output defines the following quantities:
\[
\begin{aligned}
&a_0(n_{xy},n) = (n-n_{xy})p_{10}\\
&a_1(n_{xy},n) = -(n_{xy}+1)(1-p_{10}-p_{01})\\
&a_2(n_{xy},n) = -(n+n_{xy}+2)p_{10} \\
&R(j,k,n_{xy}) = - \frac{j(n+n_{xy}+2)p_{10}}{n_{xy}+j+1},\\
\end{aligned}
\]
which satisfy the recurrence relation
\begin{equation}
\label{eq:ap:rec1}
\begin{aligned}
&a_0(n_{xy},n)S(j,n_{xy},n) + a_1(n_{xy},n)S(j,n_{xy}+1,n) + \\
&a_2(n_{xy},n)S(j,n_{xy}+2,n) =\\
&R(j+1,n_{xy},n)S(j+1,n_{xy},n) - R(j,n_{xy},n)S(j,n_{xy},n).
\end{aligned}
\end{equation}
Summing Equation (\ref{eq:ap:rec1}) over $j$ gives the recurrence for $P(N_{xy}=n_{xy})$. We confirm that the right hand side is zero:
\[
\begin{aligned}
&a_0(n_{xy},n)P(N_{xy}=n_{xy}) + a_1(n_{xy},n)P(N_{xy}=n_{xy}+1) + \\
& a_2(n_{xy},n)P(N_{xy}=n_{xy}+2) =\\ 
&\sum_{j=0}^{\lfloor \frac{n-n_{xy}}{2}\rfloor}\left(R(j+1,n_{xy},n)S(j+1,n_{xy},n) - R(j,n_{xy},n)S(j,n_{xy},n)\right)=\\
&R\left(\lfloor\frac{n-n_{xy}}{2}\rfloor +1,n_{xy},n\right)
{S\left(\lfloor\frac{n-n_{xy}}{2}\rfloor+1,n_{xy},n\right)}
- \\
&
{R(0,n_{xy},n)}
S(0,n_{xy},n) =0.
\end{aligned}
\]
That $R(0,n_{xy},n)=0$ follows straight from the definition, and that $S(\lfloor\frac{n-n_{xy}}{2}\rfloor+1,n_{xy},n) = 0$ follows via $S_{j+1} = (S_{j+1}/S_{j})S_{j}$ and Equation (\ref{eq:ap:pk_sumrat}).

\subsection*{Derivation of Equation (\ref{eq:insequence})}
%========================================================

\emph{In-sequence and between-sequence correlation}
\vspace{1ex}

\label{sec:ap:overlap}

As mentioned in the main text, PSPM based annotation of motifs generates in-sequence dependence that is not per-se accounted for in our model. Suppose there is a first order (Markov) dependence of $X_i$ on $X_{i-1}$, quantified by the parameter $\lambda_x$ (and likewise for $Y_i$). Under these assumptions the expected value for $N_x$ is still $k_xp$, but for the variance we find \cite{klotz_1973_statistical}:
\begin{equation}
\label{eq:mean_var_markov}
\begin{aligned}
&\mbox{Var}(N_x) =\\
&k_xp(1-p) + \frac{2p(1-p)(\lambda_x -p)}{1-\lambda_x} \times \\
&\left[ (k_x-1) - \frac{\lambda_x -p}{1-\lambda_x} \left(1-\left[\frac{\lambda_x -p}{1-p}\right]^{k_x} \right)  \right],
\end{aligned}
\end{equation}
and an equivalent expression for $N_y$. For $N_{xy} = N_x-N_y$ we then find (assuming no between-sequence dependence)
\begin{equation}
\label{eq:var_markov}
\begin{aligned}
\mbox{Var}(N_{xy}) = 	\;\;&k_xp(1-p) + A(p,\lambda_x,k_x) +\\
			    &k_yq(1-q) + A(q,\lambda_y,k_y) \\
\end{aligned}
\end{equation}
where $A(\cdot,\cdot,\cdot)$ represents the second term in the variance formula in Equation (\ref{eq:mean_var_markov}) and $\mbox{Cov}(X_i,Y_i) = 0$. Comparing Equation (\ref{eq:var_markov}) with Equation (\ref{eq:mean_var_kdl}), substituting $p=p_{11}+p_{10}$ and $q=p_{11}+p_{01}$ we arrive at Equation (\ref{eq:insequence}) after some algebra.
Note that a negative correlation between $X_i$ and $X_{i+1}$ decreases the variance in $N_x=\sum_iX_i$, and similarly for $N_y$. If both sequences have negative correlation between subsequent successes, the variance of $N_{xy}$ decreases. This is the same effect a correlation between $X_i$ and $Y_i$ has on the variance of $N_{xy}$.  

\subsection*{Parameter estimates contributing to $\hat{\rho}$ in Equations (\ref{eq:insequence}) and (\ref{eq:betweensequence})}
%==============================================================================================================================

\emph{Count-based and model-based parameter estimates}
\vspace{1ex}

\label{sec:ap:evol}

Here we describe estimates for the parameters $\lambda_x = \lambda_y =: \lambda$ and ${p}_{1\rightarrow 1}$, which in our model quantify in-sequence and between-sequence dependencies, respectively ($\lambda=P(X_i=1|X_{i-1}=1)$ and $p_{1\rightarrow 1} = P(Y_i = 1| X_i=1)$, see Section \ref{sec:modelParameters} in the main text). 
We assume that the in-sequence dependence is the same in $x$ and $y$, that motif gains and losses are time-reversible (i.e., $P(Y_i=1|X_i=1) = P(X_i=1|Y_i=1)$ and present count-based estimates as well as estimates based on a phylogenetic hidden Markov model (phyloHMM).

\emph{Count-based estimates:$\quad$} For a count-based estimate for $\lambda$, we count the number of adjacent motif hits in both $x$ and $y$, and then divide it by the number of overall motif hits in both sequences. This is analogous to the estimate $\hat{p}$ for the success probability of the two Binomial trials $X$ and $Y$, as described in the main text. Correspondingly, for $\hat{p}_{1\rightarrow 1}$ we count the number of congruent motif hits in $x$ and $y$, and then divide the result by the overall motif hits in $x$.  The advantage of these estimates is that they do not take much effort to calculate. The downside is that typically $p$ is small (for instance because a strict Type I error cutoff $t$ in motif prediction, see Section \ref{sec:motifpred}). This, in turn, means that (especially for short and intermediate length sequences)  not many adjacent or congruent motif hits will be observed. This, in turn, can make these count-based estimates very variable in those situations. 

\emph{Model-based estimates:$\quad$} To overcome the variability in the count-based estimates described above to some extent, we assume a phyloHMM as an underlying, generative model for the two sequences $x$ and $y$. We fit this model to our observation (the sequences $x$ and $y$, plus the corresponding motif hits) and then derive the parameters of interest as large sample properties from the fitted model.
As described in the main text, the phyloHMM consists of three states: a background state (corresponding to a neutral evolutionary model), a motif state, and a state for the reverse complement of the motif. 
First, we model the transition probabilities to be $\zeta/2$ for background-to-motif and motif-to-motif transitions, and $1-\zeta$ for background-to-background and motif-to-background transitions. We then fix the $\zeta$ in such a way that 
\begin{equation}
\label{eq:hitprob}
E_\mathcal{O}[P_{S\sim\mathcal{O}}(T(S)>t)] = \hat{p},
\end{equation}

\noindent where $\hat{p}$ is our estimate of the success probability, $S$ is a nucleotide sequence of motif-length (with log odds score $T(S)$) emitted by the phyloHMM $\Psi$ as either of the two sequences, and $\mathcal{O}$ is the state-path of motif-length generated by the Markov chain in $\Psi$ that underlies the emission of $S$. Note that the LHS in Equation (\ref{eq:hitprob}) depends on $\zeta$ because the probability for each state-path depends on the transition probabilities; but the LHS  is independent of $\tau$, because the $\Psi$ is time-reversible and $S$ is a ``marginalized'' single sequence, not a sequence-pair. 
To evaluate the expectation in Equation (\ref{eq:hitprob}) we enumerate all possible state-paths $\mathcal{O}$ and calculate ($i$) their Type I motif-hit error according to the PSPM and background distribution used for motif annotation (see Section \ref{sec:modelBasic}), and ($ii$) their probability of occurrence from the equilibrium frequencies of the Markov chain. 
This yields an estimate $\hat{\zeta}$.

Next, to obtain an estimate for $\tau$ we maximize the likelihood of the sequences $x$ and $y$:
\[
\hat{\tau} = \mbox{argmax}_\tau \;L( (x,y) | \Psi(\tau,\hat{\zeta})),
\]

\noindent where $L()$ denotes the likelihood of jointly observing $x$ and $y$. Overall this procedure yields a fully specified (fitted) phyloHMM $\Psi(\hat{\tau},\hat{\zeta})$.

Finally, we use this fitted phyloHMM to obtain estimates for $\lambda$ and  $p_{1\rightarrow 1}$. To that end we generate two very long (100,000 nucleotides or longer) sequences and take ($i$) $\hat{\lambda}$ to be the fraction of adjacent motif hits, and ($ii$) $\hat{p}_{1\rightarrow 1}$ the to be the fraction of motif hits that is congruent between the two generated sequences. We note that it is straight forward to obtain PAC bounds for these estimates via Binomial tail inversion \cite{kaariainen_2005_comparison}.

\section*{Acknowledgements}
This work was supported by grants from the National Institutes of Health (\#GM82901 and \#HL098179), a National Science Foundation graduate fellowship, and institutional funds from the Gladstone Institutes and the University of Pittsburgh School of Medicine. 

\bibliography{motifDiverge}{}
\bibliographystyle{ieeetr}

\end{document}